\documentclass[onecolumn,amsmath,showkeys,amssymb]{revtex4}

\usepackage{graphicx}
\usepackage{color}

\usepackage{amsbsy,amsmath,amsfonts, amssymb}
\usepackage{amsthm,mathrsfs,amsopn}

\begin{document}

\newcommand*{\wfnthree}{\mathrel{\hspace{0.2em}\hbox{$\bullet$}\hspace{-0.2em}\hbox{$-$}\hspace{-0.2em}\hbox{$\bullet$}\hspace{-0.2em}\hbox{$-$}\hspace{-0.2em}\hbox{$\bullet$}}}

\title{Turing instabilities on Cartesian product networks}

\author{Malbor Asllani$^{1}$, Daniel M. Busiello$^{2}$, Timoteo Carletti$^3$, Duccio Fanelli$^4$ and Gwendoline Planchon$^{2,4}$}

\affiliation{1. Dipartimento di Scienza e Alta Tecnologia, University of Insubria, via Valleggio 11, 22100 Como, Italy\\
2. Department of  Physics and Astronomy G. Galilei, University of Padova, via Marzolo 8, 35131 Padova, Italy\\
3. Department of mathematics and Namur Center for Complex Systems - naXys, University of Namur, rempart de la Vierge 8, B 5000 Namur, Belgium\\
4. Dipartimento di Fisica e Astronomia, University of Florence and INFN, Via Sansone 1, 50019 Sesto Fiorentino, Florence, Italy}

\begin{abstract} 
The problem of Turing instabilities for a reaction-diffusion system
defined on a complex Cartesian product networks is considered. To this end we operate in the linear regime and expand the 
time dependent perturbation on  a basis formed by the tensor product of the eigenvectors of the discrete Laplacian operators, associated to each of the 
individual networks that build the Cartesian product. The dispersion relation which controls the onset of the instability depends on a set of discrete 
wavelenghts, the eigenvalues of the aforementioned Laplacians. Patterns can develop on the Cartesian network, if they are supported on at least one of its 
constituive sub-graphs. Multiplex networks are also obtained under specific prescriptions. In this case, the criteria for the instability reduce to compact explicit formulae.
Numerical simulations carried out for the Mimura-Murray reaction kinetics confirm the adequacy of the proposed theory.
\end{abstract}

\keywords{Cartesian product networks, Complex networks, Nonlinear dynamics, Reaction-diffusion systems, Spatio-temporal patterns, Turing patterns}
\maketitle

\vspace{0.8cm}

\section{Introduction}
\label{sec:intro}
Patterns are widespread in nature and appear in large plethora of different conformations. From chemistry to biology, 
passing through physics, beautiful spatially extended motifs are found which spontaneously emerge from 
an ensemble made of interacting microscopic actors. The spirals in chemical reactions, the colorful patterns on fish skin, and the maculated 
fur coat of felines are all examples which testify on the intrinsic ability of natural system to self-organize, both in space and in time~\cite{murray,zhab}.

The proto-typical approach to patterns formation in reaction-diffusion processes dates back to Alan Turing's seminal paper on morphogenesis~\cite{turing}. 
Working within a simplified deterministic model for two species in mutual interactions, Turing proved that a homogeneous fixed point can turn unstable to external perturbations. A symmetry breaking instability can in fact develop which is seeded by diffusion and necessitates of   
an activator-inhibitor scheme of interaction between factors. When the conditions for the Turing instability are satisfied, the perturbation grows exponentially at short times and the system evolves towards an asymptotic stationary stable attractor, characterized by a
spatially inhomogeneous density distribution. Mathematical conditions for the onset of the instability can be obtained via a linear stability analysis, which requires expanding the imposed perturbation on the complete basis formed 
by the eigenvectors of the Laplacian operators on the chosen domain. Turing instabilities are usually studied on regular lattices or continuous supports. The theory of patterns formation extends however to reaction-diffusion systems defined on a complex graph,
as illustrated in the pioneering paper by Othmer and Scriven~\cite{othmer}, and recently revisited by Nakao and Mikhailov~\cite{nakao}. In this case the domain of the dispersion relation, from which the instability conditions ultimately descend, is the spectrum of the discrete Laplacian associated to the embedding network. Laplacian eigenvalues determine in fact the spatial characteristics of the emerging patterns, when the system is defined on a heterogeneous complex support. Turing patterns for systems defined on a complex graphs materialize in a segregation into activator-rich and activator-poor nodes \cite{Malbor2014}. As discussed in~\cite{PRE2014}, self-organized patterns can also manifest on multiplex, networks of networks assembled as adjacent layers \cite{mucha,gomez,bianconi,morris,nicosia,kivela,boccaletti,massaro}. 
Remarkably, patterns on a multiplex can be instigated by a constructive interference between layers, also when the Turing-like instability is 
prevented to occur on each single layer taken separately. In other cases, inter-layer diffusion can instead act a destructive pressure on the process of 
pattern formation~\cite{PRE2014}.

Building on these premises, we here aim at generalizing the theory of Turing instability for reaction diffusion systems defined on 
Cartesian networks. These latter are assembled as the Cartesian product of simpler networks, the fundamental 
building blocks in the process of hierarchical aggregation. Regular grids, cubes, and their counterparts in higher dimensions are for instance obtained from the Cartesian product of linear chains. Besides the interest from a graph theory point of view~\cite{Vizing1963,Imrich2000}, Cartesian product (also referred to as Cartesian networks in the following) have have been recently used in the framework of control processes~\cite{chapman2012} and systems synchronisation~\cite{atay2005}.

In this paper we shall adapt the linear instability analysis to the relevant setting of the Cartesian networks, and elaborate on the condition for the instability, by expanding the perturbation on a generalized basis formed 
by the tensor product of the eigenvectors of the discrete Laplacian operators, defined on each individual network. For a sake of clarity we will illustrate the theory with reference to the simplified setting where the Cartesian product involves two distinct networks. Clearly, one can straightforwardly extend the theory to Cartesian products made by more than two networks. The generalized process of patterns formation on the Cartesian support will be thoroughly discussed in conjunction with the standard analysis which applies to each of the graphs taken independently. We will then consider the special case of multiplex networks assembled via the Cartesian product procedure and prove that the patterns can be created or destroyed by adding more layers to the structure.

The paper is organized as follows. In the next Section we will present the general theory of Cartesian product networks and formulate the problem of patterns formation for reaction--diffusion systems defined on such networks. For a generic choice of the diffusivities, we shall prove that patterns emerge in the Cartesian product provided they can develop in at least one of the two networks from which the Cartesian support originates. In this respect, 
Cartesian products are more prone to exhibit Turing instabilities than their corresponding factor networks. In the limiting case when the diffusivities do not depend 
on the topology of the networks, but just on the species ability to relocate to neighbors sites, Turing patterns can set in if and only if the instability takes place on both factor networks.  Our analytical conclusions will be challenged numerically by 
employing the Mimura--Murray model \cite{MimuraMurray1978} as a representative reaction scheme.  We will then turn to investigate the conditions for the emergence of self-organized  patterns on degenerate multiplex networks - the same network is repeated on all layers - an important case study which can be handled as an immediate byproduct of our analysis. Finally, in the last Section, we will sum up and conclude.

\section{Results}
\label{sec:results}

\begin{figure}[htbp!]
\centering
\includegraphics[width=17cm]{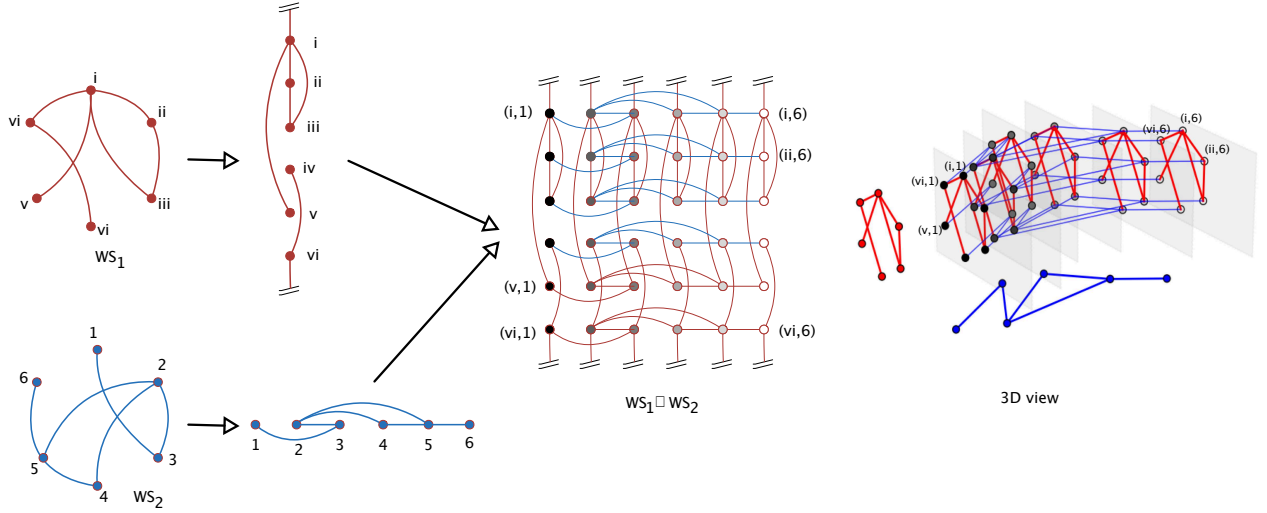}
\caption{{\bf The Cartesian product of two Watts-Strogatz~\cite{WattsStrogatz} $1$D networks}. Left panel: the two Watts-Strogatz networks $WS_1$ (red on line) and $WS_2$ (blue on line) are presented by using respectively a circular and a linear layout. This latter yields the planar representation of the Cartesian Product network, as displayed in the middle panel. According to this picture the Cartesian network appears as a \lq\lq perturbation\rq\rq of a regular 2D lattice. Right panel: the $3D$  view of the Cartesian Product. Links in the Cartesian Product are coloured as in the individual factor network (the same convention has been used in the middle panel). Nodes are instead coloured according to a  grey scale for each layer of the product, from white (corresponding to node $6$ in $WS_2$) to black (corresponding to node $1$ in $WS_2$). The cylindrical shape is obtained by imposing a circular layout to $WS_1$ and a chain layout to $WS_2$.}
\label{fig:wsxws}
\end{figure}

Given two networks $G$ and $H$, being respectively characterized by $n_G$ and $n_H$ nodes, hereby denoted $g_i\in V_G$ and $h_j\in V_H$, and by edges 
$(g_i,g_j)\in E_G$ and $(h_i,h_j)\in E_H$, one can build~\cite{Vizing1963,Imrich2000} their {\em Cartesian product} $G\square H$, that is the network 
composed by $n_G n_H$ nodes $V_G\times V_H$ and whose edges $E_{G\square H}$ are defined by:
\begin{equation}
\label{eq:edgesCP}
e=((g_1,h_1),(g_2,h_2))\in E_{G\square H} \text{ iff $g_1=g_2$ and $(h_1,h_2)\in E_H$ or $h_1=h_2$ and $(g_1,g_2)\in E_G$.} 
\end{equation}
We represent in Fig.~\ref{fig:wsxws} an example of a Cartesian product network built from two Watts-Strogatz networks~\cite{WattsStrogatz}.

Let $A^G$, respectively $A^H$, be the adjacency matrix of the network $G$, respectively $H$. Then the adjacency matrix of the Cartesian product network $G\square H$ is given by
\begin{equation}
\label{eq:adjCP}
A^{G\square H}= A^G\otimes \mathbb{I}_{n_H}+\mathbb{I}_{n_G}\otimes A^H\, ,
\end{equation}
where $\mathbb{I}_{n}$ is the $n\times n$ identity matrix and $ \otimes$ is the Kronecker product. Let us recall that the Kronecker product of two matrices $A$ and $B$, is the matrix 
\begin{equation*}
A\otimes B=\left(\begin{matrix}a_{11}B & \hdots & a_{1n}B\\ \vdots & \ddots & \vdots\\ a_{n1}B & \hdots & a_{nn}B\end{matrix}\right)\, ,
\end{equation*}
where $a_{ij}$ are the elements of matrix $A$.
Let $L^G=A^G-D^G$ (respectively $L^H=A^H-D^H$) be the Laplacian matrix of the network $G$ (respectively of $H$), where $D^{G}$ is the diagonal matrix whose entries are the degrees of the network $G$ (similarly for $D^H$). Then the Laplacian matrix of the Cartesian product network $G\square H$ reads:
\begin{equation}
\label{eq:lapCP}
L^{G\square H}=L^G\otimes \mathbb{I}_{n_H}+\mathbb{I}_{n_G}\otimes L^H\, .
\end{equation}
From the latter equation, it straightforwardly follows that the eigenvalues of $L^{G\square H}$ are of the form
\begin{equation}
\label{eq:eigvalCP}
\Lambda^{G\square H}_{\alpha \beta}=\Lambda^{G}_{\alpha}+\Lambda^{H}_{\beta}\quad \forall \,{\alpha}=1,\dots n_G \text{ and } \,{\beta}=1,\dots n_H\, ,
\end{equation}
that is the eigenvalues of the Laplacian matrix associated to the Cartesian product are obtained as the sum of the eigenvalues of the Laplacian operators of each factor 
networks. Moreover, the eigenvectors of $L^{G\square H}$ are given by $\phi_{\alpha\beta}=(\phi_{\alpha}^G,\phi_{\beta}^H)$, where
$\phi_{\alpha}^G$ and $\phi_{\alpha}^H$ stand for the eigenvectors of respectively $L^G$ and $L^H$. In fact:
\begin{eqnarray*}
L^{G\square H}(\phi_{\alpha}^G,\phi_{\beta}^H)&=&(L^G\otimes \mathbb{I}_{n_H})(\phi_{\alpha}^G,\phi_{\beta}^H)+(\mathbb{I}_{n_G}\otimes L^H)(\phi_{\alpha}^G,\phi_{\beta}^H)\\
&=&(L^G\phi_{\alpha}^G,\phi_{\beta}^H)+(\phi_{\alpha}^G,L^H\phi_{\beta}^H)=(\Lambda^{G}_{\alpha}\phi_{\alpha}^G,\phi_{\beta}^H)+(\phi_{\alpha}^G,\Lambda^{H}_{\beta}\phi_{\beta}^H)\\
&=&\Lambda^{G}_{\alpha}(\phi_{\alpha}^G,\phi_{\beta}^H)+\Lambda^{H}_{\beta}(\phi_{\alpha}^G,\phi_{\beta}^H)=(\Lambda^{G}_{\alpha}+\Lambda^{H}_{\beta})(\phi_{\alpha}^G,\phi_{\beta}^H)\, .
\end{eqnarray*}
Let us observe that $L^G$ and $L^H$ are zero sum symmetric and negative--semidefinite matrices and so it is $L^{G\square H}$. Hence, the eigenvalues 
$L^{G\square H}$ are all negative, except for the largest one which is identical to zero. We will organize the list eigenvalues so that the first position ($\alpha=\beta=1$) reads always zero, the largest eigenvalue. Hence, $\Lambda_1^{G\square H}=0$.

\vspace{0.5truecm}

{\bf Reaction-Diffusion systems on Cartesian product networks}
%\label{sec:rdcp}
\\
Let us now consider a reaction--diffusion system defined on a Cartesian product network $G\square H$. To this end we introduce two species whose continuous 
density are labelled $u$ and $v$. The two species undergo local interaction when they share one of the $n_Gn_H$ nodes of $G\square H$ and diffuse among adjacent sites
via existing links. Denote with $D^G_u\geq 0$, respectively $D^G_v\geq 0$, the diffusion coefficient of species $u$, respectively $v$, on network $G$. 
For network $H$, one can introduce the homologous quantities $D^H_u\geq 0$ and $D^H_v\geq 0$. In the following we shall indicate with $u_{gh}$ and $v_{gh}$ 
the concentrations of respectively $u$ and $v$ at node $(gh)\in G\square H$. As usual, local rules of interaction among species translate in non linear functions of the concentration amount, hereafter $f(u_{gh},v_{gh})$ and $g(u_{gh},v_{gh})$. The diffusion is in turn modeled by resorting to conventional Laplacian operators. In formulae:

\begin{equation}
\label{eq:evolnet}
\begin{cases}
\dot{u}_{gh}&= f(u_{gh},v_{gh})+ \mathcal{L}_u u_{gh} \\
\dot{v}_{gh}&= g(u_{gh},v_{gh})+ \mathcal{L}_v v_{gh} \, ,
\end{cases}\quad \forall g\in \{1,\dots,n_G\}\, ,h\in \{1,\dots,n_H\}\text{ and $t>0$}\, .
\end{equation}
where the diffusion operator $\mathcal{L}_s$ (with $s=u,v$) reads: 

\begin{equation}
\label{eq:diffoper}
\mathcal{L}_s=D_s^G L^G\otimes \mathbb{I}_{n_H}+D^H_s\mathbb{I}_{n_G}\otimes L^H\quad \text{for $s=u,v$}\, .
\end{equation}

Notice that $u_{gh}$ can be written as $(u^G_{g},u^H_{h})$, and hence  $(L^G\otimes \mathbb{I}_{n_H})(u^G_{g},u^H_{h})=(L^Gu^G_{g},u^H_{h})$. On the other hand, 
$L^Gu^G_{g}=\sum_{g^{\prime}} L^G_{gg^{\prime}}u^G_{g^{\prime}}$, which implies $(L^Gu^G_{g},u^H_{h})=(\sum_{g^{\prime}} L^G_{gg^{\prime}}u^G_{g^{\prime}},u^H_{h})=\sum_{g^{\prime}} L^G_{gg^{\prime}}u_{g^{\prime}h}$. 
Similar considerations hold for $\mathbb{I}_{n_G}\otimes L^H$, and  one can therefore rewrite~\eqref{eq:evolnet} as:

\begin{equation}
\label{eq:evolnet2}
\begin{cases}
\dot{u}_{gh}&= f(u_{gh},v_{gh})+D_u^G (L^G u)_{gh}+D_u^H (L^H u)_{gh}=f(u_{gh},v_{gh})+D_u^G \sum_{g^{\prime}}L^G_{gg^{\prime}} u_{g^{\prime}h}+D_u^H \sum_{h^{\prime}}L^H_{h^{\prime}h}u_{gh^{\prime}} \\
\dot{v}_{gh}&= g(u_{gh},v_{gh})+D_v^G (L^G v)_{gh}+D_v^H (L^H v)_{gh}=g(u_{gh},v_{gh})+D_v^G \sum_{g^{\prime}}L^G_{gg^{\prime}} v_{g^{\prime}h}+D_v^H \sum_{h^{\prime}}L^H_{h^{\prime}h}v_{gh^{\prime}} \, .
\end{cases}
\end{equation}

To progress in the analysis we shall assume that an homogeneous solution of the above equations exists, 
i.e. $({u}_{gh},{v}_{gh})=(\hat{u},\hat{v})$, for all $g$ and $h$  such that  $f(\hat{u},\hat{v})=g(\hat{u},\hat{v})=0$. In addition,  we will require the homogeneous 
fixed point $(\hat{u},\hat{v})$ to be stable, which in turn amounts to impose $\mathrm{tr}(J)=\partial_u f+\partial_vg<0$ and 
$\det(J)= \partial_u f\partial_vg-\partial_v f\partial_ug>0$, where $J$ stands for the Jacobian matrix evaluated at $(\hat{u},\hat{v})$ (to keep the notation simple and because $f$ and $g$ do not depend on the nodes index, we have replaced $u_{gh}$ and $v_{gh}$ by $u$ and $v$ in the former and in their derivatives). Following the standard Turing 
recipe, we set down to study the  conditions that yields an exponential growth of a non-homogeneous perturbation around $(\hat{u},\hat{v})$.
We hence define $\delta u_{gh}=u_{gh}-\hat{u}$ and $\delta{v}_{gh}={v}_{gh}-\hat{v}$ and linearize system 
(\ref{eq:evolnet2}) around the equilibrium
\begin{equation}
\label{eq:evolnet_lin}
\begin{cases}
\dot{\delta u}_{gh}&= f_u {\delta u}_{gh} +f_v{\delta v}_{gh}+D_u^G \sum_{g^{\prime}}L^G_{gg^{\prime}}\delta u_{g^{\prime}h}+D_u^H \sum_{h^{\prime}}L^H_{h^{\prime}h}\delta  u_{gh^{\prime}} \\
\dot{\delta v}_{gh}&= g_u\delta u_{gh}+g_v\delta v_{gh}+D_v^G \sum_{g^{\prime}}L^G_{gg^{\prime}} \delta v_{g^{\prime}h}+D_v^H \sum_{h^{\prime}}L^H_{h^{\prime}h}\delta v_{gh^{\prime}} \, ,
\end{cases}
\end{equation}
where $f_u$, $f_v$, $g_u$ and $g_v$ are the derivatives of $f$ and $g$ with respect to $u$ and $v$ evaluated at the equilibrium point $(\hat{u},\hat{v})$.

To go one step further we expand $\delta u_{gh}$ and $\delta{v}_{gh}$ on the eigenbasis of the Laplacian matrix for $G\square H$ and look for solution of  
system (\ref{eq:evolnet_lin}) in the form:
\begin{equation*}
\delta u_{gh}=\sum_{\alpha \beta} U_{\alpha \beta}\phi_{\alpha \beta}^{g h}e^{\lambda_{\alpha \beta}t}\quad \text{and}\quad \delta v_{gh}=\sum_{\alpha \beta} V_{\alpha \beta}\phi_{\alpha \beta}^{g h}e^{\lambda_{\alpha \beta}t}\, .
\end{equation*}
By inserting the previous relations into the linearized system (\ref{eq:evolnet_lin}), one readily finds that the following condition should be met for a 
non-trivial solution to exist: 
\begin{equation*}
\det\left(\tilde{J}-\lambda_{\alpha \beta}\mathbb{I}\right)=0\, ,
\end{equation*}
where 
\begin{equation*}
\tilde{J}=\left(\begin{matrix} f_u+D_u^{G}\Lambda^{G}_{\alpha}+D_u^{H}\Lambda^{H}_{\beta} & f_v\\
g_u &g_v+D_v^{G}\Lambda^{G}_{\alpha}+D_v^{H}\Lambda^{H}_{\beta}\end{matrix}\right)
\, ,
\end{equation*}
that is
\begin{equation}
\label{2ord_eq}
\lambda_{\alpha,\beta}^2-P(\Lambda^{G}_{\alpha},\Lambda^{H}_{\beta})\lambda_{\alpha,\beta}+Q(\Lambda^{G}_{\alpha},\Lambda^{H}_{\beta})=0\, ,
\end{equation}
where $P(\Lambda^{G}_{\alpha},\Lambda^{H}_{\beta})$ and $Q(\Lambda^{G}_{\alpha},\Lambda^{H}_{\beta})$ are defined as:

\begin{eqnarray}
\label{eq:FHvh}
P(\Lambda^{G}_{\alpha},\Lambda^{H}_{\beta})&=&\mathrm{tr}(J)+\Lambda^{H}_{\beta}(D_u^{H}+D_v^{H})+\Lambda^{G}_{\alpha}(D_u^{G}+D_v^{G})\\
Q(\Lambda^{G}_{\alpha},\Lambda^{H}_{\beta})&=&\det(J)+\Lambda^{H}_{\beta}(D_u^{H}g_v+D_v^{H}f_u)+\Lambda^{G}_{\alpha}(D_u^{G}g_v+D_v^{G}f_u)\notag\\&+&(\Lambda^{H}_{\beta})^2D_u^{H}D_v^{H}+\Lambda^{H}_{\beta}\Lambda^{G}_{\alpha}(D_u^{H}D_v^{G}+D_u^{G}D_v^{H})+(\Lambda^{G}_{\alpha})^2D_u^{G}D_v^{G}\notag\, .
\end{eqnarray}

Let us observe that $P$ is always negative because of the stability assumption ($\mathrm{tr}(J)<0$) and since
$\Lambda^{H}_{\beta}, \Lambda^{G}_{\alpha}\leq 0$. The exponential instability manifests provided the real part of $\lambda_{\alpha,\beta}$ gets positive over a bounded portion of 
the plane $(\Lambda^{H}_{\beta}, \Lambda^{G}_{\alpha})$. For this reason, we shall solely concentrate on the largest root of equation (\ref{2ord_eq}): 
 
\begin{equation*}
\lambda^{G\square H}(\Lambda^{G}_{\alpha},\Lambda^{H}_{\beta})=\frac{P(\Lambda^{G}_{\alpha},\Lambda^{H}_{\beta})+ \sqrt{P^2(\Lambda^{G}_{\alpha},\Lambda^{H}_{\beta})-4Q(\Lambda^{G}_{\alpha},\Lambda^{H}_{\beta})}}{2}\, .
\end{equation*}

whose real part is also called the dispersion relation. Turing instability develops on the Cartesian network provided 
\begin{equation}
\label{eq:Hneg}
Q(\Lambda^{G}_{\alpha},\Lambda^{H}_{\beta})<0\, .
\end{equation}
within a finite domain  in  $\Lambda^{H}_{\beta}$ and $\Lambda^{G}_{\alpha}$. In the following we shall set
$\lambda^{G\square H}(\Lambda^{G}_{\alpha},\Lambda^{H}_{\beta})\equiv \lambda_{\alpha,\beta}(\Lambda^{G}_{\alpha},\Lambda^{H}_{\beta})$ and $Q^{G\square H}(\Lambda^{G}_{\alpha},\Lambda^{H}_{\beta}) \equiv Q(\Lambda^{G}_{\alpha},\Lambda^{H}_{\beta})$  to make explicit 
reference to the embedding Cartesian topology. 

The above derivation can be formally adapted to the simpler case where the 
reaction-diffusion system is defined on a standard graph $G$. In this case the condition for the existence of Turing patterns amounts to imposing $Q^G(\Lambda^{G}_{\alpha})=\det(J)+\Lambda^{G}_{\alpha}(D_u^{G}g_v+D_v^{G}f_u)+(\Lambda^{G}_{\alpha})^2D_u^{G}D_v^{G}<0$, inside a bounded interval of $\Lambda_{\alpha}^G$. Importantly, $Q^G(\Lambda^{G}_{\alpha})=Q^{G\square H}(\Lambda^{G}_{\alpha},0)$. Similar considerations hold for graph $H$, which is combined to $G$ to yield the Cartesian network ${G\square H}$. In practical terms, the 
dispersion relation which controls the instability on a Cartesian support is a multi-dimensional function (two dimensional, for the case under exam), which reduces to the conventional one dimensional function, when projected on each of the independent subspaces that compose the Cartesian backing. Notice that the above conclusions can be also reached by employing a straightforward two dimensional extension  
of the network-targeted Fourier transform introduced in \cite{Malbor2012,Malbor2013} to the current multi-dimensional setting. 

Starting from this scenario, it is interesting to elaborate on the mathematical conditions
that underly the emergence of collective patterns on a  Cartesian support, in relation to the mechanisms which 
seed the homologous instabilities on the composing graphs, taken separately. Are Cartesian patterns  reminiscent of the instability that occur on each layer of the assembly? To answer this question it is entirely devoted the remaining part of the paper.

\vspace{0.5truecm}

{\bf Different diffusion constants on distinct graphs.} 
\\
Let us start by considering the general case where the diffusion coefficients for each species are assumed to depend on the hosting network, namely
$D^G_u\ne D^H_u$ and $D^G_v \ne D^H_v$. Imagine that Turing patterns can develop when the inspected reaction-diffusion system is hosted on $G$. Then, as we shall prove hereafter, the patterns can invade the Cartesian support $G\square H$. Similar conclusions obviously hold when the dual scenario is considered, i.e. when the patterns are allowed to develop on graph $H$, instead of $G$. 

Since Turing patterns can be found by hypothesis on network $G$, there exists at least one $\hat{\alpha}\in \{1,\dots, n_G\}$ such that $Q^G(\Lambda_{\hat{\alpha}}^G)<0$. Consider the eigenvalue $\Lambda^{G\square H}_{\hat{\alpha} 1}=\Lambda^{G}_{\hat{\alpha}}+\Lambda^{H}_{1}=\Lambda^{G}_{\hat{\alpha}}$ (where in the last step we made use of $\Lambda^{H}_{1}=1$) 
and write the following chain of relations:
\begin{equation*}
Q^{G\square H}(\Lambda^G_{\hat{\alpha}},\Lambda^H_1)\equiv Q^{G\square H}(\Lambda^G_{\hat{\alpha}},0)=Q^G(\Lambda_{\hat{\alpha}}^G)<0\, .
\end{equation*}
The same modes which are unstable on $G$, are also destabilized when the reaction-diffusion system is made to evolve on the Cartesian support $G\square H$. The network $G\square H$ can hence exhibit Turing patterns, the perturbation being localized on a set of unstable modes which includes (or coincide with) those active on $G$.  If the spectrum of the Laplacian was continuum, one could always delimit, by continuity of $Q^{G\square H}$, a finite portion of the parameter plan  $(\Lambda^{G},\Lambda^{H})$, adjacent to the degenerate line  $(\Lambda^G,0)$, for which $Q^{G\square H}<0$. 
However, the spectrum  of the Laplacian operator is discrete. One should therefore require 
a sufficiently small $\lvert \Lambda_{\bar{\beta}}^H\rvert $ to exist, so that $Q^{G\square H}(\Lambda^G_{\hat{\alpha}},\Lambda^H_{\bar{\beta}})<0$, 
for non trivial modes of the Cartesian support could be triggered  unstable.

%Assume a small enough, positive $\epsilon$ exists such that $\lvert \Lambda_{\bar{\beta}}^H\rvert <\epsilon$. Then one can excite non trovial modes in $G\square H$ provided:
%\begin{equation*}
%Q^{G\square H}(\Lambda^G_{\hat{\alpha}},\Lambda^H_{\bar{\beta}})\leq Q^G(\Lambda^G_{\hat{\alpha}})+\epsilon c\, ,
%\end{equation*}
%for some positive constant $c$, thus $Q^{G\square H}(\Lambda^G_{\hat{\alpha}},\Lambda^H_{\bar{\beta}})$ is negative if $\epsilon$ is small enough.

\begin{figure}[htbp!]
\begin{tabular}{ccc}
\includegraphics[width=6cm]{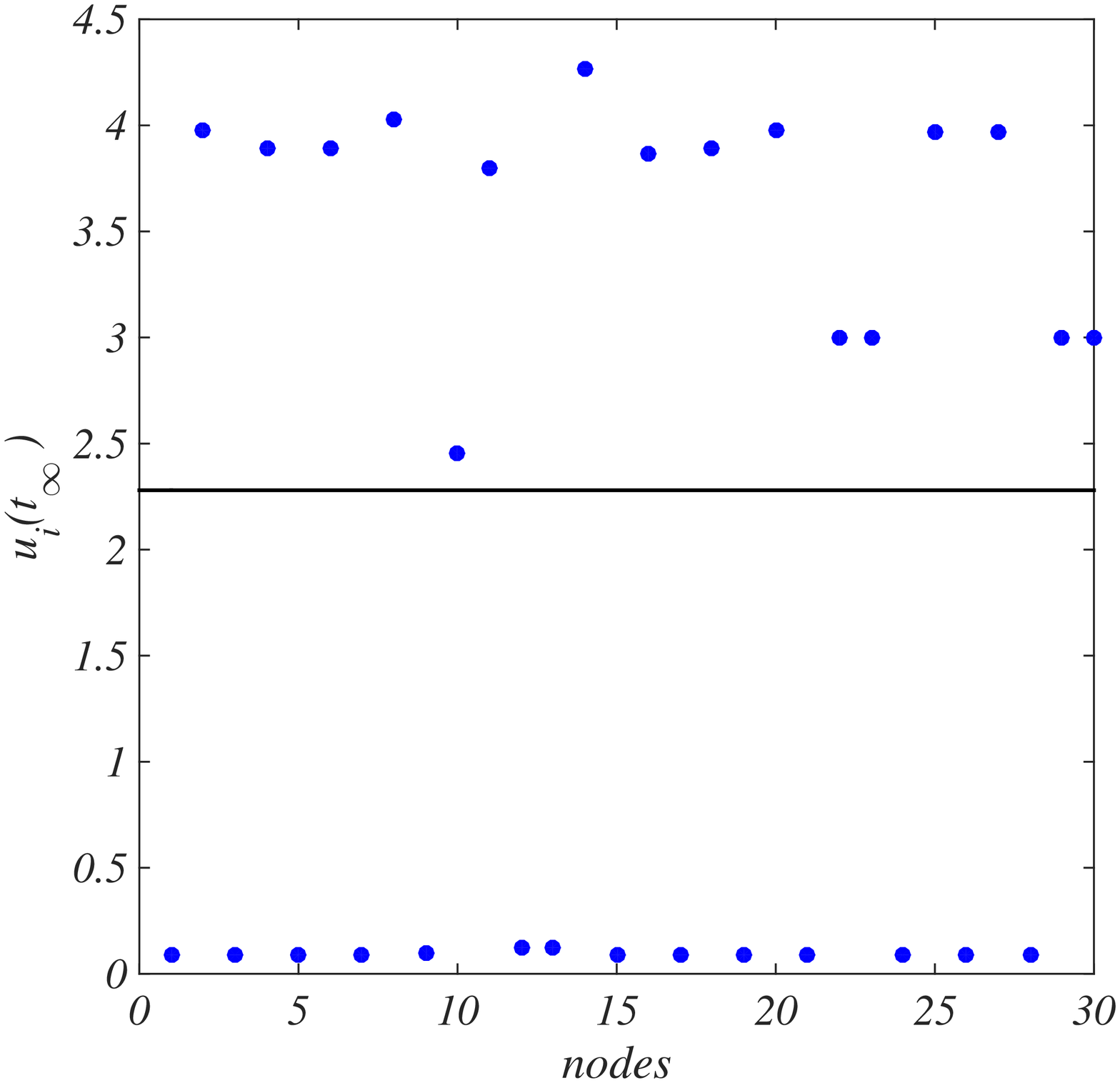}&
\includegraphics[width=6cm]{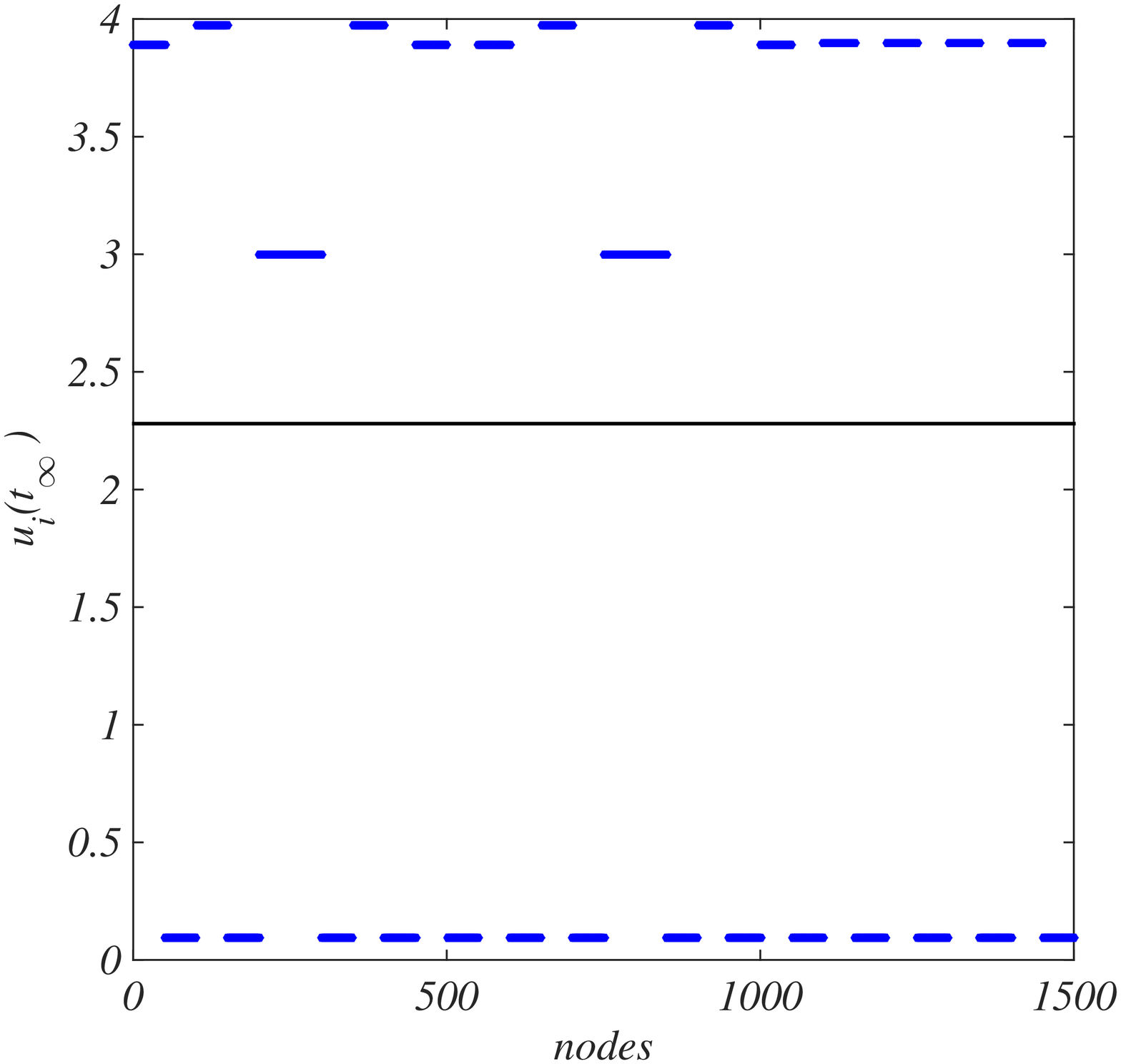}&
\includegraphics[width=6cm]{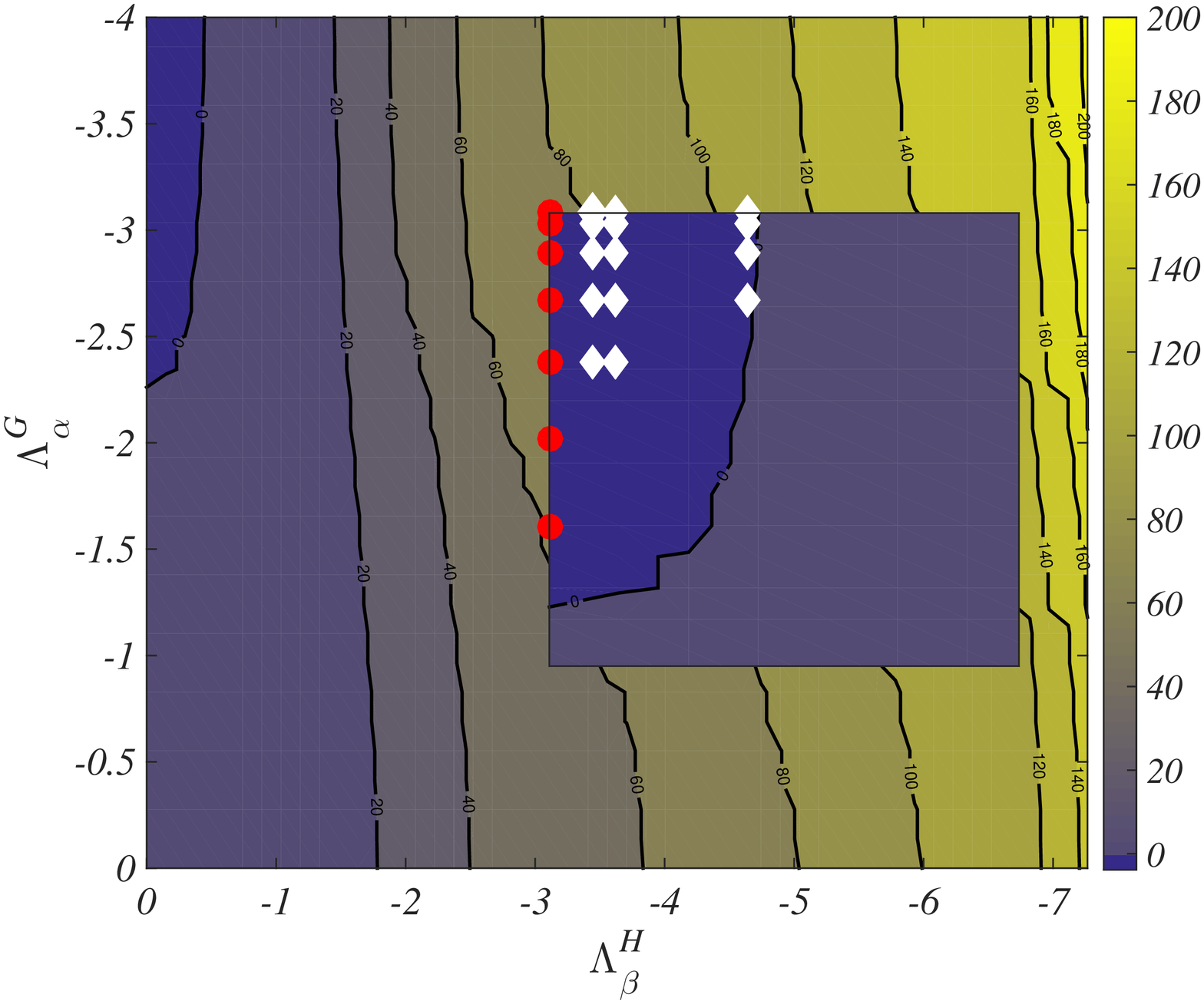}\\
(a)& (b)&(c)
\end{tabular}
\caption{{\bf Turing patterns for the Mimura-Murray model on the Cartesian product $G\square H$.} Panel (a): $G$ exhibits Turing patterns, the asymptotic concentration of species $u$ (circles, blue online) varies from node to node and differs from the correpsonding equilibrium value $\hat{u}\sim 2.28$ (horizontal line). The diffusion constants are set so that Turing patterns cannot develop on graph $H$ (data not shown). Panel (b): the Mimura-Murray model defined on the cartesian product $G\square H$ displays Turing patterns. Once again the asymptotic concentration of species $u$ (circles, blue online)  varies from node to node and differs from the predicted equilibrium solution (horizontal black line). Panel (c): level sets of the function $Q_{\alpha\beta}^{G\square H}$. Notice the (dark blue online) zone in the top left corner of the figure (enlarged in the inset) where the function takes negative values. The ensemble of vertical (red online) circles represents the eigenvalues of $L^G$. As anticipated, $Q^{G\square H}(\Lambda^G_{\alpha},0)=Q^{G}(\Lambda^G_{\alpha})<0$, which points to the existence of Turing instability on subspace $G$. White diamonds identify the pairs $(\Lambda^G_{\alpha},\Lambda^H_{\beta})$ for which $Q^{G\square H}(\Lambda^G_{\alpha},\Lambda^H_{\beta})<0$: patterns are hence supported on the Cartesian product network $G\square H$. On the other hand,  for $\Lambda^G_{\alpha}=0$, the function $Q^{G\square H}$ is positive. One cannot find $\Lambda^H_{\beta}$ for which $Q^{G\square H}(0,\Lambda^H_{\beta})<0$, in agreement with our initial working assumption: patterns cannot grow on graph $H$, when taken isolated. Here, $G$ and $H$ are Watts-Strogatz~\cite{WattsStrogatz} networks composed respectively of $n_G=30$ and $n_H=50$ nodes. Their associated links rewiring probabilities are taken to $p_G=0.01$ and $p_H=0.04$ and the average degree are given by $<k_G>=2$ and $<k_G>=4$. The diffusion coefficients are set to the values $D_u^G = 0.01$, $D_v^G = 2.1$, $D_u^H = 1.12$ and $D_v^H = 2.6$. The initial condition is a perturbation of the homogeneous fixed point ($\hat{u},\hat{v}$). Such externally imposed perturbation is node dependent, hence inhomogeneous, drawn from a uniform distribution and scaled with an  amplitude factor $\delta=0.005$.}
\label{fig:GyesHnoGHyes1}
\end{figure}

To make this concept more explicit, we consider the celebrated Mimura-Murray model~\cite{MimuraMurray1978}, which we shall assume to specify the reaction terms. More specifically we will set  $f(u,v)= \left( (a+bu-u^2)/c - v\right) u$ and $g(u,v)= \left( u - (1+dv) \right) v$, where $a, b, c$ and $d$ are constant parameters.
The Mimura-Murray model possesses six equilibria, whose stability depends on the value of the above parameters. We will hereby set  
$a=35$, $b=16$, $c=9$ and $d=0.4$ and focus on the homogeneous stationary solution $\hat{u}= 1+(bd - 2d - c + \sqrt{\Delta})/(2d)$, $\hat{v}= (bd - 2d - c + \sqrt{\Delta})/(2d^2)$ where $\Delta = (bd-2d-c)^2 + 4d^2(a + b -1)$. It is immediate to realize that $\det(J)>0$ and $\mathrm{tr}(J)<0$, hence the selected fixed point is stable. The diffusion coefficients are assigned as discussed in the caption of Fig.~\ref{fig:GyesHnoGHyes1}. In particular,
patterns can develop when the Mimura-Murray system is let evolve on graph $G$. At variance, Turing instability cannot take place on graph $H$. When the system is instead hosted on the Cartesian support $G\square H$, as obtained by composing together the individual graphs $G$ and $H$, patterns can materialize, as 
demonstrated in Fig. \ref{fig:GyesHnoGHyes1}. The generalized dispersion relation takes indeed positive values over a finite portion of the discrete two dimensional support $(\Lambda^G_{\alpha},\Lambda^H_{\beta})$ as it can be appreciated by visual inspection of panel c of Fig. \ref{fig:GyesHnoGHyes1}.

\vspace{0.5truecm}

{\bf The diffusion is the same on distinct networks} 
\\
Consider now the simpler setting where the diffusion coefficients are assumed identical on all graphs composing the Cartesian networks. In formulae we require $D^G_u=D^H_u=D_u$ and $D^G_v=D^H_v=D_v$. Also the kinetics parameters do not depend on the reaction site. Under this working hypothesis, Turing patterns are allowed on the Cartesian network $G\square H$,  if and only if they can also develop on both $G$ ot $H$. To prove our clain, we remark that the assumption of identical diffusivities enables one to simplify the dispersion relation and  Eqs.~\eqref{eq:FHvh} and, in particular, we get:
\begin{equation}
\label{eq:FHvhsamediff}
Q(\Lambda^{G}_{\alpha},\Lambda^{H}_{\beta})=\det(J)+(\Lambda^{G}_{\alpha}+\Lambda^{H}_{\beta})(D_ug_v+D_vf_u)+D_uD_v(\Lambda^{H}_{\beta}+\Lambda^{G}_{\alpha})^2\notag\, .
\end{equation}
Hence,  the instability takes place on the Cartesian support if $D_ug_v+D_vf_u>0$ and 
$(D_ug_v+D_vf_u)^2-4\det(J)D_uD_v>0$. On the other hand, when these latter inequalities are matched, Turing patterns develop on both $G$ and $H$, provided their discrete Laplacian eigenvalues populate the interval where the dispersion one dimensional dispersion relations  
$\lambda_{\alpha}^G$ and $\lambda_{\alpha}^H$ are positive. 

\vspace{0.5truecm}

\begin{figure}[htbp!]
\centering
\includegraphics[width=10cm]{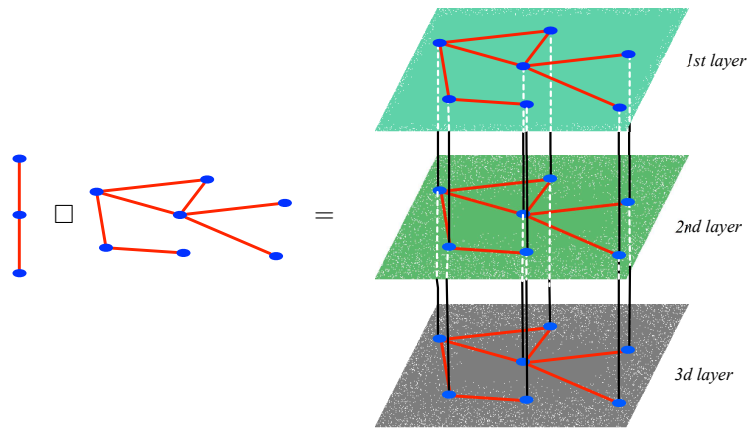}
\caption{{\bf The degenerate multiplex.} The Cartesian product of $G=\wfnthree$ and a generic graph $H$.}
\label{fig:cpchain}
\end{figure}

{\bf Degenerate multiplex as the Cartesian product of two graphs.} 
\\
Assume $G$ to be an open one dimensional chain, with nearest neighbors connections. This  configuration is also termed {\em path} in the literature, and  differs from a ring or cycle, because it lacks periodic boundary conditions. Then, for any arbitrary choice of network $H$, the cartesian product $G\square H$ is a multiplex with peculiar characteristics.  On each layer of the multiplex the same network $H$ is repeated. Inter-layer connections are established only between adjacent layers, as depicted in Fig.~\ref{fig:cpchain}.

If we are interested in investigating the possibility of a Turing like instability for a generic reaction-diffusion system defined on such a multiplex, one cannot resort to the approach discussed in~\cite{PRE2014}. The multiplex is in fact degenerate, meaning that the layers are identical by construction and their associated spectra 
coincide. This clearly implies dealing with repeated eigenvalues a condition which violates the hypothesis on which the analysis of ~\cite{PRE2014} builds \footnote{It is however worth emphasising that the analysis of ~\cite{PRE2014} can be extended to the case where the eigenvalues are indeed repeated, at the price of some additional complications in the calculations}. Following the above conclusion, we can however expect Turing patterns to materialize on the multiplex support, if the reaction-diffusion system under inspection can undergo a diffusion driven instability when placed on the path network $G$. As we shall argue in the following, this request translates in a compact condition for the instability to develop on the multiplex support. In fact, the homogenous equilibrium is unstable to external inhomogeneous perturbation, for a reaction diffusion-system evolving on $G$, provided    
 $f_u D^v_G+g_vD^u_G>0$ and $(f_u D^v_G+g_vD^u_G)^2-4\det(J)D^u_GD^v_G>0$. On the other hand, the eigenvalues of the Laplacian operator defined on $G$ can be written in a closed form as:

\begin{equation}
\label{eq:eiglinch}
\Lambda^G_{\alpha}=-2+2\cos\frac{{\alpha}-1}{n_G}\pi\quad \alpha\in\{1,\dots, n_G\}\, .
\end{equation}

\begin{figure}[htbp!]
\begin{tabular}{cc}
\includegraphics[width=7cm]{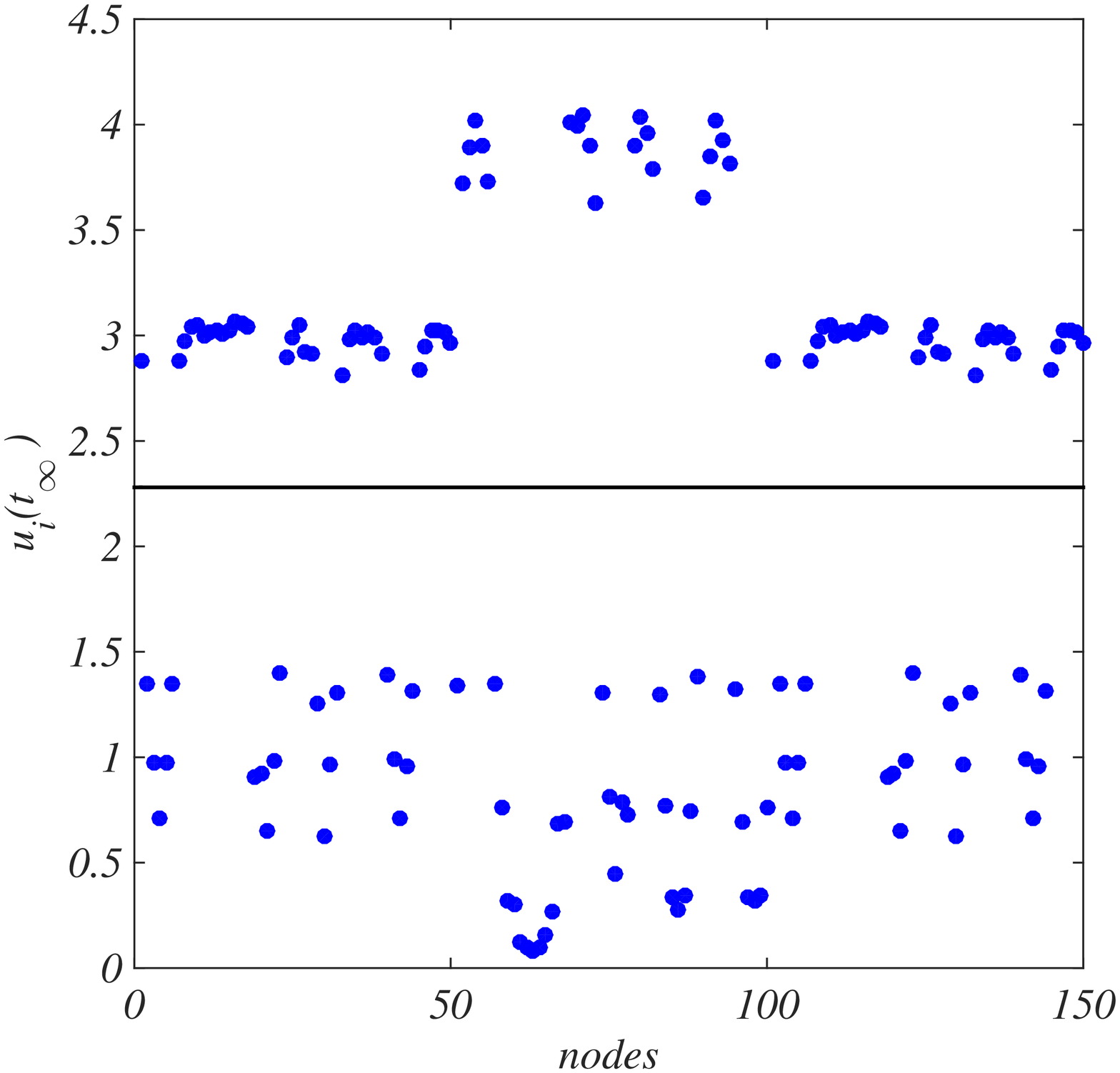}&
\includegraphics[width=7cm]{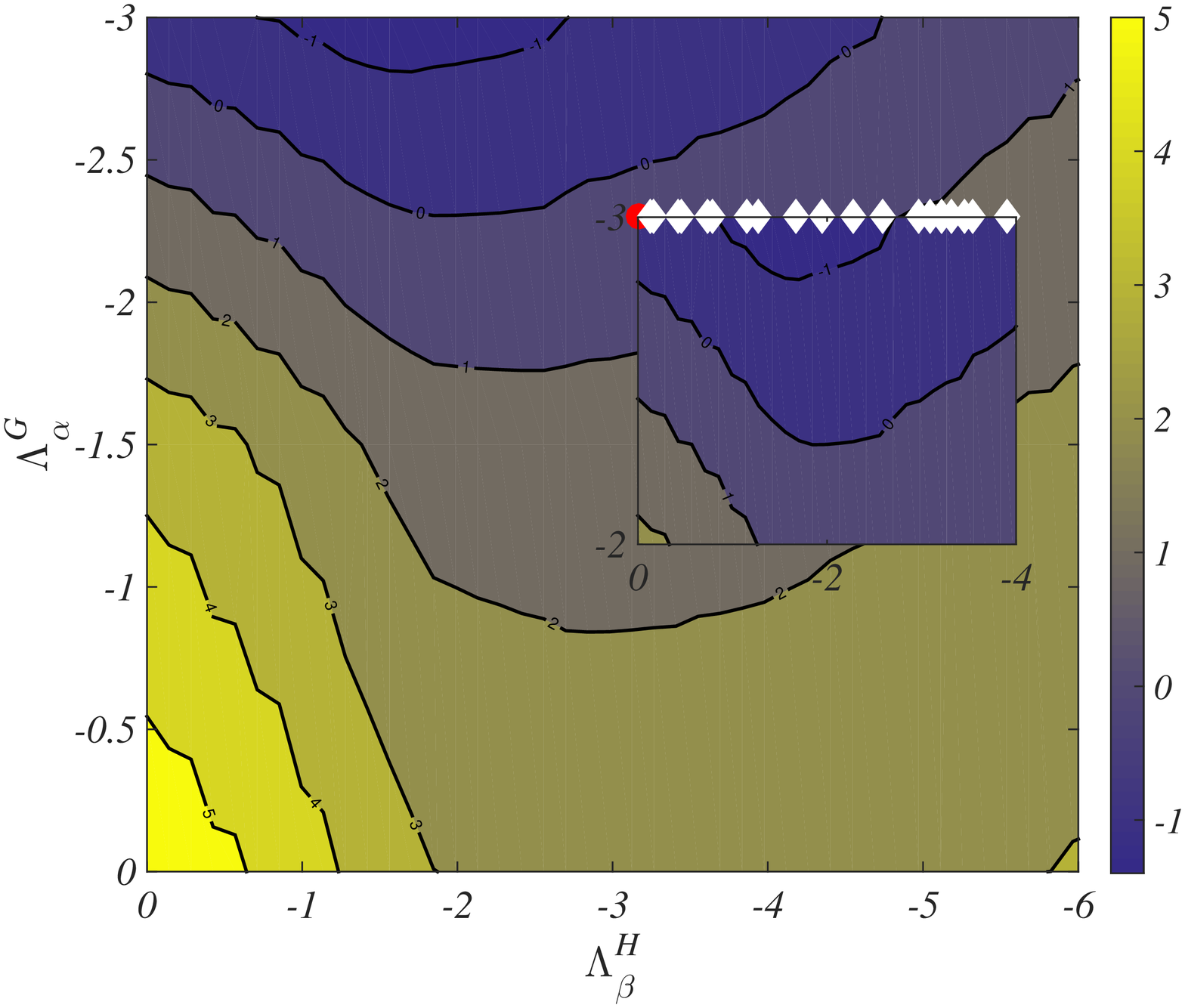}\\
(a)& (b)
\end{tabular}
\caption{{\bf Turing patterns for the Mimura-Murray model on a degenerate multiplex support.} The multiplex is built as the Cartesian product $G\square H$. Here, $G=\wfnthree$ (namely, it is a linear chai with nearest neighbors connections) and $H$ is a Watts-Strogatz~\cite{WattsStrogatz} network composed by $n_H=50$ nodes, with a probability to rewire a link equal to $p_H=0.04$ and average degree $<k_H>=4$. Panel (a): asymptotic distribution for the concentration of species $u$ (circles, blue online) on each node of the Cartesian product $G\square H$.  The recorded concentration varies from node to node and 
differs from the deputed equilibrium value $\hat{u}\sim 2.28$ (horizontal black line). For the selected parameters (see below), the Mimura-Murray 
model is Turing unstable on the linear chain $G$. Turing patterns cannot develop instead, when the reaction-diffusion model is made to evolve on the Watts--Strogatz network $H$ alone. Panel (b): level sets of the function $Q_{\alpha\beta}^{G\square H}$. In the top border of the picture (region enlarged in the inset, dark blue online) the function assumes negative values. The (red online) circle identifies the unstable eigenvalue of the Laplacian operator associated to the linear chain. As anticipated, it falls in the region $Q^{G\square H}(\Lambda^G_{\alpha},0)=Q^{G}(\Lambda^G_{\alpha})<0$, hence signalling the presence of Turing-like patterns on $G$. The white diamonds refer to the pairs $(\Lambda_{\alpha}^G,\Lambda_{\beta}^H)$ for which $Q^{G\square H}(\Lambda^G_{\alpha},\Lambda_{\beta}^H)<0$, thus implying the existence of the instability on the Cartesian product $G\square H$. We remark that for $\Lambda^G_{\alpha}=0$, the function $Q^{G\square H}$ is positive defined: patterns cannot emerge when the diffusion of the interacting species is  
confined on graph $H$. Here, the diffusion coefficients are set to the representative values $D_u^G = 0.005$, $D_v^G = 1.8$, $D_u^H = 0.12$ and $D_v^H = 1.3$. The initial condition is assigned as explained in the caption of Fig.~\ref{fig:GyesHnoGHyes1}.}
\label{fig:multiplexCP}
\end{figure}

For a given reaction kinetics, Turing patterns can flourish on $G$, if and only if there exists at least on eigenvalue $\Lambda^G_{\hat{\alpha}}$ for which $Q^G(\Lambda^G_{\hat{\alpha}})<0$, namely
\begin{equation}
\label{cond_multiplex}
q_-<\Lambda^G_{\hat{\alpha}}<q_+\, ,
\end{equation}
where
\begin{equation}
q_{\pm}=\frac{-(f_u D^v_G+g_vD^u_G)\pm\sqrt{(f_u D^v_G+g_vD^u_G)^2-4\det(J)D^u_GD^v_G}}{2D^u_GD^v_G}\, ,
\end{equation}
are the two positive roots of $Q^G_{\hat{\alpha}}=0$. When condition \eqref{cond_multiplex} is met, and by virtue of the analysis carried out above,
the patterns can invade the multiplex support. In Fig. \ref{fig:multiplexCP} we provide a direct evidence of the phenomenon, employing again the Mimura-Murray reaction model as the reference case study. 

\begin{figure}[htbp!]
\begin{tabular}{ccc}
\includegraphics[width=5.5cm]{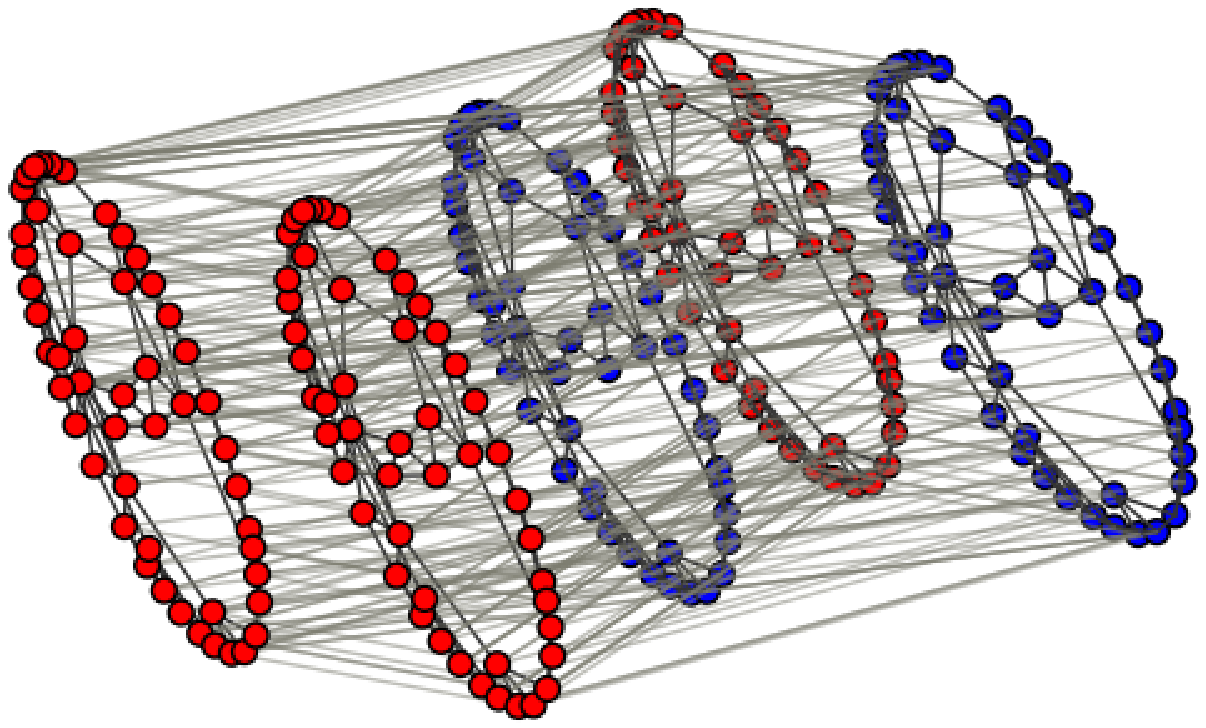}&
\includegraphics[width=7cm]{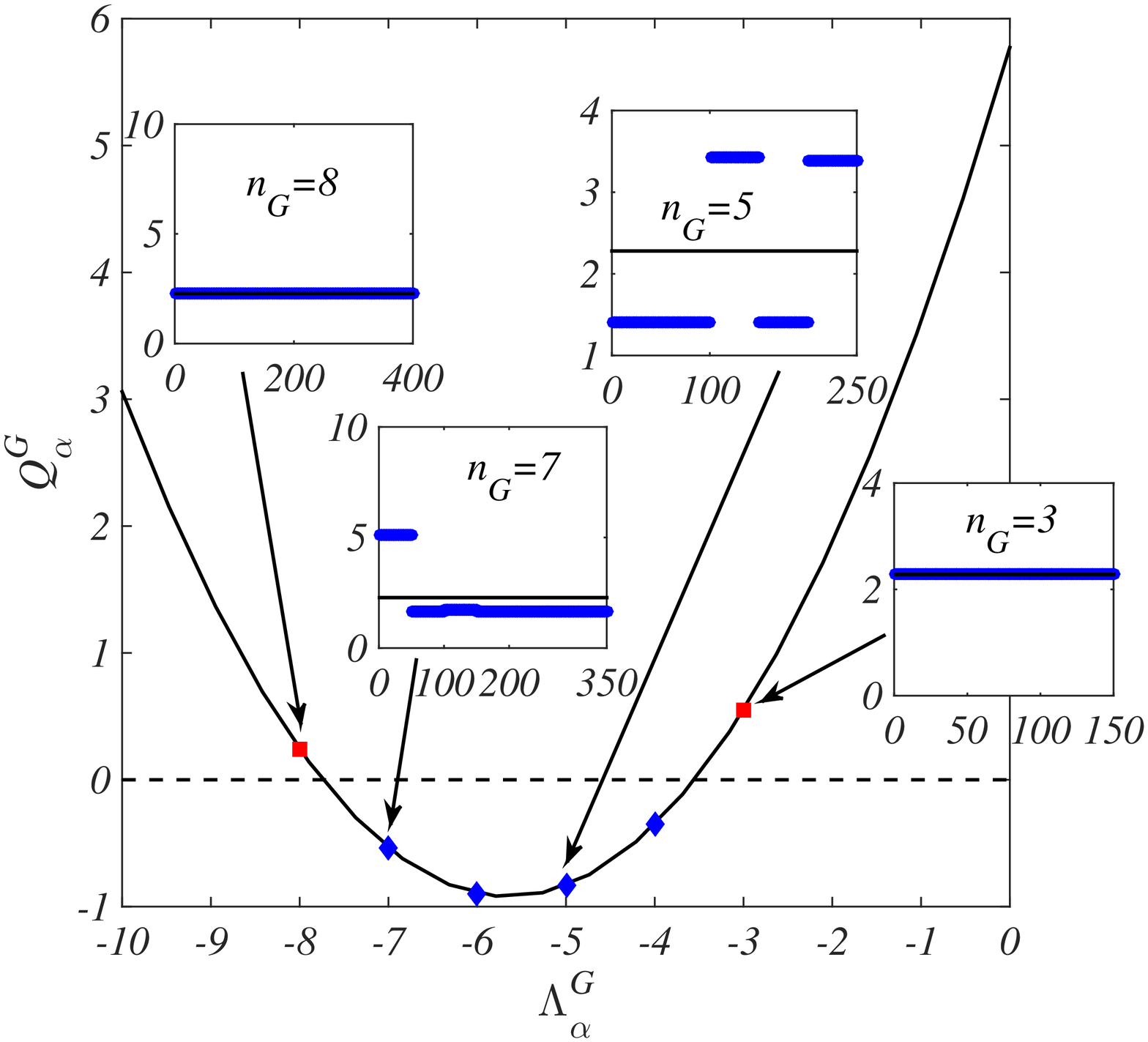}&
\includegraphics[width=5.5cm]{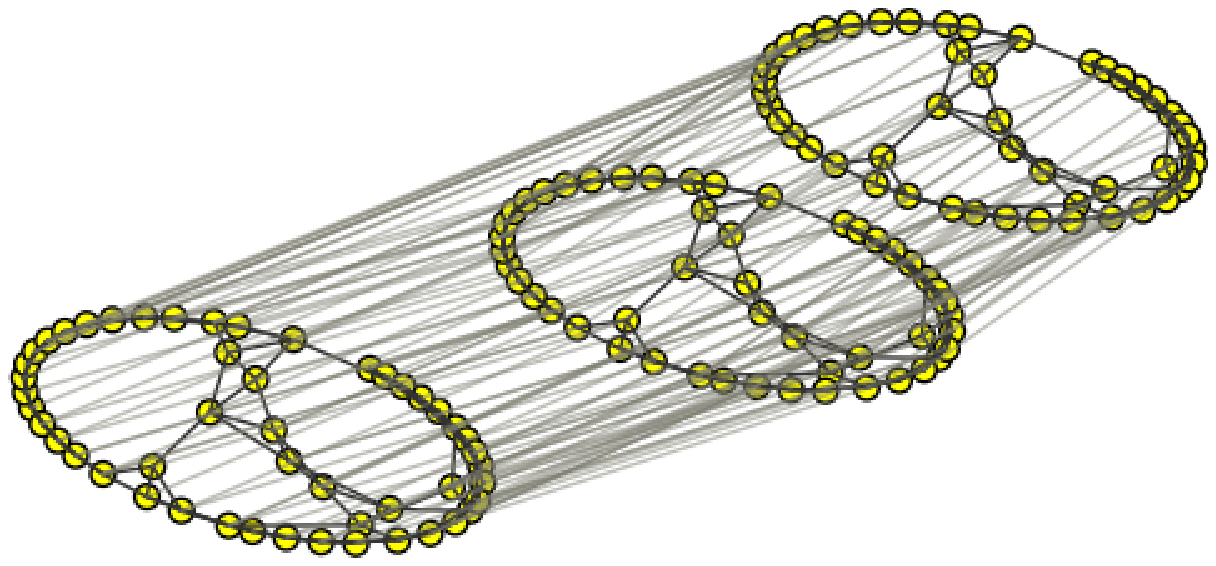}\\
$K_5\square H$ & & $K_3\square H$
\end{tabular}
\caption{{\bf Turing patterns for the Mimura-Murray model on a multiplex $G\square H$, with $G$ complete network.} Left panel: Turing Patterns (for the species $u$) in the multiplex $K_5\square H$ ($n_G=5$), where $H$ is the network obtained by using the Watts-Strogatz algorithm~\cite{WattsStrogatz} with $n_H=50$, $p_H=0.04$ and average degree $<k_H>=4$. Blue nodes correspond to nodes where the asymptotic concentration of species $u$ is larger than the homogeneous values $\hat{u}\sim 2.28$ ($u_i(t_{\infty})-\hat{u}>\hat{u}/10$); red nodes correspond to nodes where the asymptotic concentration of species $u$ is lower than the homogeneous values $\hat{u}\sim 2.28$ ($u_i(t_{\infty})-\hat{u}<-\hat{u}/10$). To help the reader only $40\%$ of links among different layers have been drawn. Notice that the patterns results in a segregation between activator rich - activator poor layers. This is an interesting self-organized stationary solution of the reaction diffusion model, that we will describe in details elsewhere. Middle panel: $Q^G(\Lambda_{\alpha}^G)$ is plotted versus $\Lambda_\alpha^G$. $n_G$ denotes the nodes of the complete network $G$. Once again, $H$ is a Watts-Strogatz~\cite{WattsStrogatz} network composed by $n_H=50$ nodes, with a probability to rewire a link equal to $p_H=0.04$ and average degree $<k_H>=4$. Turing patterns can develop if and only if  $Q^G(\Lambda_{\alpha}^G)$ for some $\Lambda_\alpha^G$. For the complete network $\Lambda_1^G=0$ and $\Lambda_\alpha^G=-n_G$ for $\alpha>1$ with multiplicity $n_G-1$. Hence, $Q_{\alpha}^G$ is negative if and only if $q_-<-n_G<q_+$. For our choice of the parameters (see below), one finds $q_-\sim -7.74$ and $q_+\sim -3.55$. Turing patterns can hence develop on $G$, and thus on the multiplex $G\square H$, if and only if $-7\leq n_G\leq -4$. This is confirmed by inspection of the annexed insets, where the asymptotic concentration of species $u$ is reported against an integer which runs over the nodes, for different choices of $n_G$. This result follows a numerical integration of the relevant reaction-diffusion equations.  
 The horizontal solid lines represent the unperturbed homogeneous fixed point. Here, the diffusion coefficients are $D_u^G = 0.1$, $D_v^G = 2.1$, $D_u^H = 1.12$ and $D_v^H = 2.6$. The initial condition is set as explained in the caption of Fig.~\ref{fig:GyesHnoGHyes1}. Right panel: absence of Turing Patterns (for the species $u$) in the multiplex $K_3\square H$ ($n_G=3$), where $H$ is again a network obtained using the Watts-Strogatz algorithm with $n_H=50$, $p_H=0.04$ and average degree $<k_H>=4$. Yellow nodes correspond to nodes where the asymptotic concentration of species $u$ is equal to the homogeneous values, $u_i(t_{\infty})=\hat{u}\sim 2.28$, for all $i$. To help the reader only $40\%$ of links among different layers have been drawn.}
\label{fig:multiplexCPcomplete}
\end{figure}

Another interesting case to consider is when $G$ is the complete graph with $n_G$ nodes, namely a network with all-to-all connections but self--loops. Then, for any network $H$, the Cartesian product $G\square H$ is a multiplex, which hosts on every layer a replica of $H$, each node of a given layer being directly connected to all its specular {\it images} on the other layers. The number of nodes of network $G$ determines therefore the number of layers of the Cartesian multiplex. Based on the above, we can readily infer an explicit condition for the existence of Turing instability on the generalized Cartesian support. The only non trivial eigenvalue of the complete graph is $-n_G$ (with multiplicity $n_G-1$) and condition \eqref{cond_multiplex} yields $q_-<-n_G<q_+$. In other words, the number of nodes of $G$, or equivalently the number of layers in the multiplex, can act as a control parameter to instigate, or alternatively dissolve, the Turing instability. In Fig.  
(\ref{fig:multiplexCPcomplete}) we provide a numerical demonstration of the predicted phenomenon. Patterns can be seen on the Cartesian multiplex, for a given choice of $H$ and of the reaction kinetics, only if the number of nodes of the complete graph $G$ falls within a bounded interval.  

\section{Conclusions}
\label{sec:conc}  

Reaction-diffusion systems on complex networks are gaining attention because of their multifaceded applications to a vast realm of interdisciplinary problems. 
As follows a symmetry breaking instability, seeded by diffusion, a stable homogeneous fixed point of the examined reaction kinetics can turn unstable to  inhomogeneous perturbation. This event is the precursor of a Turing pattern, which is eventually approached by the system in its late time evolution. The conditions that underly the spontaneous emergence of self-organized patterns can be obtained via a standard linear stability analysis, which requires expanding the imposed perturbation on a basis formed by the eigenvectors of the network Laplacian operator. Importantly, the associated eigenvalues play the role of non trivial wavelenghts for the embedding network support. 

Starting from this setting, we have here extended the analysis to Cartesian networks, namely generalized complex support
assembled as the Cartesian product of simpler networks, and elaborated on the conditions for the Turing instability to set in. 
To this aim, we introduced and exploited a generalized basis for tracking the perturbation in its linear regime of evolution. This is 
the tensor product of the eigenvectors of the discrete Laplacian operators, defined on each individual network. For simplicity we focused from the beginning on the simplified setting where the Cartesian product involves two distinct networks, but the analysis, as well as the conclusions of our study, apply to a more general setting where several networks can be combined together to give a multidimensional Cartesian Product. 
The dispersion relation which ultimately determines the onset of the instability is now function of two independent set of discrete wavelenghts, the eigenvalues 
of the Laplacian operators constructed from the two networks that combine in the Cartesian structure. As a consequence, the process of patterns formation for a reaction-diffusion system on Cartesian support can be rationalized via an integrated approach which moves from the analysis of the instability conditions on each of the graphs taken independently. In particular, we could prove that patterns can invade the Cartesian network, if they are supported on one of the graphs that compose its structure. Multiplex networks can be also obtained as a special limiting case and the domain instability delimited by compact relations. When a generic network 
is assembled with a complete graph to yield a degenerate multi-dimensional complex lattice, the onset of the instability can be controlled by the number of nodes of the complete sub-structure. Our findings  have been corroborated by direct numerical integration of the reaction-diffusion equations, assuming the Mimura-Murray kinetics as a representative model.

\section*{Acknowledgments}
The work of T.C. presents research results of the Belgian Network DYSCO (Dynamical Systems, Control, and Optimization), funded by the Interuniversity Attraction Poles Programme, initiated by the Belgian State, Science Policy Office.

\end{document}